\documentstyle[draft,psfig]{mn} 

\newif\ifAMStwofonts

\ifoldfss
  \ifCUPmtlplainloaded \else
    \NewTextAlphabet{textbfit} {cmbxti10} {}
    \NewTextAlphabet{textbfss} {cmssbx10} {}
    \NewMathAlphabet{mathbfit} {cmbxti10} {} 
    \NewMathAlphabet{mathbfss} {cmssbx10} {} 
  \fi
  \ifAMStwofonts
    \ifCUPmtlplainloaded \else
      \NewSymbolFont{upmath} {eurm10}
      \NewSymbolFont{AMSa} {msam10}
      \NewMathSymbol{\upi}     {0}{upmath}{19}
      \NewMathSymbol{\umu}     {0}{upmath}{16}
      \NewMathSymbol{\upartial}{0}{upmath}{40}
      \NewMathSymbol{\leqslant}{3}{AMSa}{36}
      \NewMathSymbol{\geqslant}{3}{AMSa}{3E}

       \let\ge=\geqslant
    \fi
  \fi
\fi 

\ifnfssone
  \newmathalphabet{\mathit}
  \addtoversion{normal}{\mathit}{cmr}{m}{it}
  \addtoversion{bold}{\mathit}{cmr}{bx}{it}
  \newmathalphabet{\mathbfit} 
  \addtoversion{normal}{\mathbfit}{cmr}{bx}{it}
  \addtoversion{bold}{\mathbfit}{cmr}{bx}{it}
  \newmathalphabet{\mathbfss} 
  \addtoversion{normal}{\mathbfss}{cmss}{bx}{n}
  \addtoversion{bold}{\mathbfss}{cmss}{bx}{n}
  \ifAMStwofonts
    \ifCUPmtlplainloaded \else
      \UseAMStwoboldmath
      \makeatletter
      \new@mathgroup\upmath@group
      \define@mathgroup\mv@normal\upmath@group{eur}{m}{n}
      \define@mathgroup\mv@bold\upmath@group{eur}{b}{n}
      \edef\UPM{\hexnumber\upmath@group}
      \new@mathgroup\amsa@group
      \define@mathgroup\mv@normal\amsa@group{msa}{m}{n}
      \define@mathgroup\mv@bold\amsa@group{msa}{m}{n}
      \edef\AMSa{\hexnumber\amsa@group}
      \makeatother
      \mathchardef\upi="0\UPM19
      \mathchardef\umu="0\UPM16
      \mathchardef\upartial="0\UPM40
      \mathchardef\leqslant="3\AMSa36
      \mathchardef\geqslant="3\AMSa3E

       \let\ge=\geqslant
    \fi
  \fi
\fi 

\ifnfsstwo
  \DeclareMathAlphabet{\mathbfit}{OT1}{cmr}{bx}{it}
  \SetMathAlphabet\mathbfit{bold}{OT1}{cmr}{bx}{it}
  \DeclareMathAlphabet{\mathbfss}{OT1}{cmss}{bx}{n}
  \SetMathAlphabet\mathbfss{bold}{OT1}{cmss}{bx}{n}
  \ifAMStwofonts
    \ifCUPmtlplainloaded \else
      \DeclareSymbolFont{UPM}{U}{eur}{m}{n}
      \SetSymbolFont{UPM}{bold}{U}{eur}{b}{n}
      \DeclareSymbolFont{AMSa}{U}{msa}{m}{n}
      \DeclareMathSymbol{\upi}{0}{UPM}{"19}
      \DeclareMathSymbol{\umu}{0}{UPM}{"16}
      \DeclareMathSymbol{\upartial}{0}{UPM}{"40}
      \DeclareMathSymbol{\leqslant}{3}{AMSa}{"36}
      \DeclareMathSymbol{\geqslant}{3}{AMSa}{"3E}

       \let\ge=\geqslant
    \fi
  \fi
\fi 

\ifCUPmtlplainloaded \else
  \ifAMStwofonts \else 
    \def\upi{\pi}
    \def\umu{\mu}
    \def\upartial{\partial}
  \fi
\fi

\title[Black hole--neutron star coalescence]{Newtonian hydrodynamics
of the coalescence of black holes with neutron stars II: Tidally
locked binaries with a soft equation of state.}

\author[W. H. Lee and W. Klu\'{z}niak] 
{William H. Lee$^{1,2}$ and W\l odzimierz Klu\'{z}niak$^{2,3}$ \\
$^1$ Instituto de Astronom\'{\i}a, Universidad Nacional Aut\'{o}noma
de M\'{e}xico, Apdo. Postal 70--264, Cd. Universitaria, 04510
M\'{e}xico D.F.\\
$^2$ Physics Dept., University of Wisconsin, 1150 University Ave.,
Madison, WI, 53706, USA \\
$^3$ Copernicus Astronomical Centre, ul. Bartycka 18, 00--716
Warszawa, Poland
}

\pagerange{\pageref{firstpage}--\pageref{lastpage}}
\pubyear{1998}

\begin{document}

\maketitle

\label{firstpage}

\begin{abstract}
We present a numerical study of the hydrodynamics in the final stages
of inspiral of a black hole--neutron star binary, when the binary
separation becomes comparable to the stellar radius. We use a
Newtonian three--dimensional Smooth Particle Hydrodynamics (SPH) code,
and model the neutron star with a soft (adiabatic index $\Gamma=5/3$)
polytropic equation of state and the black hole as a Newtonian point
mass which accretes matter via an absorbing boundary at the
Schwarzschild radius. Our initial conditions correspond to tidally
locked binaries in equilibrium, and we have explored configurations
with different values of the mass ratio $q=M_{\rm NS}/M_{\rm BH}$,
ranging from $q=1$ to $q=0.1$. The dynamical evolution is followed for
approximately 23~ms, and in every case studied here we find that the
neutron star is tidally disrupted on a dynamical timescale, forming a
dense torus around the black hole that contains a few tenths of a
solar mass. A nearly baryon--free axis is present in the system
throughout the coalescence, and only modest beaming of a fireball that
could give rise to a gamma--ray burst would be sufficient to avoid
excessive baryon contamination. We find that some mass (on the order
of $10^{-3}-10^{-2}$~M$_{\odot}$) may be dynamically ejected from the
system, and could thus contribute substantially to the amount of
observed r--process material in the galaxy. We calculate the
gravitational radiation waveforms and luminosity emitted during the
coalescence in the quadrupole approximation.
\end{abstract}

\begin{keywords}
binaries: close --- gamma rays: bursts --- hydrodynamics --- stars: neutron
\end{keywords}

\section{Introduction}

Angular momentum losses to gravitational radiation are expected to
lead to the coalescence of binary systems containing black holes,
and/or neutron stars (when the initial binary separation is small
enough for the decay to take place in less than the Hubble time). This
type of evolution has been suggested in a variety of contexts as
possibly giving rise to observable events, such as gamma---ray bursts
(GRBs) and bursts of gravitational waves (see
e.g. Thorne~1995). Additionally, it could help explain the observed
abundances of heavy elements in our galaxy~(Lattimer \& Schramm~1974;
1976) if the star is tidally disrupted in the encounter (see
Wheeler~1971). Study of such events could also provide constraints on
the equation of state at supra--nuclear densities.

After the recent measurement of redshifts to their
afterglows~\cite{metzger,kulkarni,djorgovski}, it is now generally
believed that GRBs originate at cosmological distances. The calculated
event rates~\cite{latt,nara,tutukov,lipunov,portyun,tomek} for merging
compact binaries are compatible with the observed frequency of GRBs
(on the order of one per day). The preferred model for the production
of a GRB invokes a relativistic fireball from a compact `central
engine' that would produce the observable $\gamma$--rays through
internal shocks~(M\'{e}sz\'{a}ros \& Rees~1992; 1993). This model
requires the presence of a relatively baryon--free line of sight from
the central engine to the observer along which the fireball can expand
at ultrarelativistic speeds. Additionally, the short--timescale
variations seen in many bursts (often in the millisecond range)
probably arise within the central engine~\cite{sari}. 

The coalescence of binary neutron star systems or black hole--neutron
star binaries was suggested as a mechanism capable of powering the
gamma--ray bursts, either during the binary merger itself or through
the formation of a dense accretion disk which could survive long
enough to accomodate the variable timescales of GRBs~(Paczy\'{n}ski
1986; Goodman 1986; Goodman, Dar \& Nussinov 1987; Eichler et
al. 1988; Paczy\'{n}ski 1991; M\'{e}sz\'{a}ros \& Rees 1992; Woosley
1993; Jaroszy\'{n}ski 1993; 1996; Witt et al. 1994; Wilson, Mathews \&
Marronetti 1996; Ruffert, Janka \& Sch\"{a}fer 1996; Lee \&
Klu\'{z}niak 1997; Ruffert, Janka, Takahashi \& Sh\"{a}fer 1997; Katz
1997; Klu\'{z}niak \& Lee 1998; Ruffert \& Janka 1998; Popham, Woosley
\& Fryer 1998; McFadden \& Woosley 1998; Ruffert \& Janka 1999). The
enormous amount of gravitational energy that would be liberated in
such an event could account for the energetics of the observed GRBs
and neutrino--antineutrino annihilation may power the necessary
relativistic fireball.

In previous work~(Lee \& Klu\'{z}niak 1995; 1998 (hereafter Paper~I);
Klu\'{z}niak \& Lee 1998), we have studied the coalescence of a
neutron star with a stellar--mass black hole for a stiff ($\Gamma=3$)
polytropic equation of state and a range of mass ratios. We found that
the neutron star was not entirely disrupted, but rather remained in
orbit (with a greatly reduced mass) about the black hole after a quick
episode of mass transfer. Thus the duration of the coalescence process
would be extended from a few milliseconds to possibly several tens of
milliseconds. The observed outcome seemed favorable for the production
of a GRB since in every case we found a baryon--free axis in the
system, along the axis of rotation.

In the present paper, we investigate the coalesence of a black
hole--neutron star binary for a soft equation of state (with an
adiabatic index $\Gamma=5/3$) and a range of mass ratios. Our initial
conditions are as in Paper~I in that they correspond to tidally locked
binaries. Complete tidal locking is not realistically
expected~\cite{bildsten}, but it can be considered as an extreme case
of angular momentum distribution in the system. In the future we will
explore configurations with varying degrees of tidal locking.

As before, the questions motivating our study are: Is the neutron star
tidally disrupted by the black hole and does an accretion torus form
around the black hole? If so, how long--lived is it? Is the baryon
contamination low enough to allow the formation of a relativistic
fireball? Is any significant amount of mass dynamically ejected from
the system? What is the gravitational radiation signal like, and how
does it depend on the equation of state and the initial mass ratio?

In section~\ref{method} we present the method we have used to carry
out our simulations. This is followed by a presentation of our results
in section~\ref{results} and a discussion in section~\ref{discussion}.

\section{Numerical method}\label{method}

For the simulations presented in this paper, we have used the method
known as Smooth Particle Hydrodynamics (SPH). Our code is
three--dimensional and essentially Newtonian. This method has been
described often, we refer the reader to
Monaghan~\shortcite{monaghan92} for a review of the principles of SPH,
and to Paper~I and Lee~\shortcite{phd} for a detailed description of
our own code, including the tree structure used to compute the
gravitational field.

We model the neutron star via a polytropic equation of state, $P=K
\rho^{\Gamma}$ with $\Gamma=5/3$. For the following, we measure
distance and mass in units of the radius {\em R} and mass $M_{\rm NS}$
of the unperturbed (spherical) neutron star (13.4~km and
1.4~M$_{\odot}$ respectively), except where noted, so that the units
of time, density and velocity are
\begin{eqnarray}
\tilde{t}= 1.146 \times 10^{-4}~{\rm s} \times
\left(\frac{R}{13.4~\mbox{km}}\right)^{3/2}
\left(\frac{M_{\rm NS}}{1.4M_{\odot}}\right)^{-1/2}
\label {eq:deftunit}
\end{eqnarray}
\begin{eqnarray}
\tilde{\rho}= 1.14 \times 10^{18}~{\rm kg~m}^{-3} \times
\left(\frac{R}{13.4~\mbox{km}}\right)^{-3}
\left(\frac{M_{\rm NS}}{1.4M_{\odot}}\right)
\label{eq:defrhounit}
\end{eqnarray}
\begin{eqnarray}
\tilde{v}=0.39c \times \left(\frac{R}{13.4~\mbox{km}}\right)^{-1/2}
\left(\frac{M_{\rm NS}}{1.4M_{\odot}}\right)^{1/2}.
\label{eq:defvunit}
\end{eqnarray}
and we use corresponding units for derivative quantities such as
energy and angular momentum.

The black hole (of mass $M_{\rm BH}$) is modeled as a Newtonian point
mass, with a potential \( \Phi_{\rm BH}(r) = -GM_{\rm BH}/r \). We
model accretion onto the black hole by placing an absorbing boundary
at the Schwarzschild radius ($r_{Sch}=2GM_{\rm BH}/c^{2}$). Any
particle that crosses this boundary is absorbed by the black hole and
removed from the simulation. The mass and position of the black hole
are continously adjusted so as to conserve total mass and total
momentum.

Initial conditions corresponding to tidally locked binaries in
equilibrium are constructed in the co--rotating frame of the binary
for a range of separations {\em r} and a given value of the mass ratio
$q=M_{\rm NS}/M_{\rm BH}$~(Rasio \& Shapiro~1994; Paper~I). The binary
separation is defined henceforth as the distance between the black
hole and the center of mass of the SPH particles. During the
construction of these configurations, the specific entropies of all
particles are maintained constant, i.e. $K$=constant in
$P$=$K\rho^{\Gamma}$. The neutron star is modeled with $N=17,256$
particles in every case presented in this paper. To ensure uniform
spatial resolution, the masses of the particles were made proportional
to the Lane--Emden densities on the initial grid.

To carry out a dynamical run, the black hole and every particle are
given the azimuthal velocity corresponding to the equilibrium value of
the angular frequency $\Omega$ in an inertial frame, with the origin
of coordinates at the center of mass of the system. Each SPH particle
is assigned a specific internal enery
$u_{i}=K\rho^{(\Gamma-1)}/(\Gamma-1)$, and the equation of state is
changed to that of an ideal gas, where $P=(\Gamma-1)\rho u$. The
specific internal energy of each particle is then evolved according to
the first law of thermodynamics, taking into account the contributions
from the artificial viscosity present in SPH. During the dynamical
runs we calculate the gravitational radiation waveforms in the
quadrupole approximation.

We have included a term in the equations of motion that simulates the
effect of gravitational radiation reaction on the components of the
binary system. Using the quadrupole approximation, the rate of energy
change for a point--mass binary is given by (see Landau \&
Lifshitz~1975):
\begin{eqnarray} 
\frac{dE}{dt}=-\frac{32}{5} \frac{G^{4}(M_{\rm NS}+M_{\rm BH})
(M_{\rm NS}M_{\rm BH})^{2}}{(cr)^{5}} 
\label{eq:dedt}
\end{eqnarray}
and the rate of angular momentum loss by
\begin{eqnarray}
\frac{dJ}{dt}=-\frac{32}{5 c^{5}} \frac{G^{7/2}}{r^{7/2}}
M_{\rm BH}^{2} M_{\rm NS}^{2} \sqrt{M_{\rm BH}+M_{\rm NS}}.
\label{eq:djdt}
\end{eqnarray} 
From these equations a radiation reaction acceleration for
each component of the binary can be obtained as
\begin{eqnarray}
\mbox{\boldmath $a$}^{*}=-\frac{1}{q(M_{\rm NS}+M_{\rm BH})} 
\frac{dE}{dt}
\frac{\mbox{\boldmath $v$}^{*}}{(v^{*})^{2}} 
\label{eq:reaction}\\
\mbox{\boldmath $a$}^{\rm BH}=-\frac{q}{M_{\rm NS}+M_{\rm BH}} \frac{dE}{dt}
\frac{\mbox{\boldmath $v$}^{\rm BH}}{(v^{\rm BH})^{2}}
\label{eq:reactionbh}
\end{eqnarray}
where $v^{*}$ is the velocity of the neutron star and $v^{\rm BH}$ that of
the black hole.

We have used this formula for our calculations to simulate the effect
of gravitational radiation reaction on the system. Clearly, the
application of equation~(\ref{eq:reactionbh}) to the black hole in our
calculations is trivial, since we always treat it as a point mass. For
the neutron star, we have chosen to apply the same acceleration to all
SPH particles. This value is that of the acceleration at the center of
mass of the SPH particles, so that equation~(\ref{eq:reaction}) now
reads:
\begin{eqnarray}
\mbox{\boldmath $a$}^{i}=-\frac{1}{q(M_{\rm NS}+M_{\rm BH})} \frac{dE}{dt}
\frac{\mbox{\boldmath $v$}^{*}_{cm}}{(v^{*}_{cm})^{2}}, 
\label{eq:reactionsph}
\end{eqnarray}
This formulation of the gravitational radiation reaction has been used
in SPH simulations by others~\cite{davies,zhuge,rosswog} in the case
of merging neutron stars, and it is usually switched off once the
stars come into contact, when the point--mass approximation clearly
breaks down. We are assuming then, that the polytrope representing the
neutron star can be considered as a point mass for the purposes of
including radiation reaction. If the neutron star is disrupted during
the encounter with the black hole, this radiation reaction must be
turned off, since our formula would no longer give meaningful
results. We have adopted a switch for this purpose, as follows: the
radiation reaction is turned off if the center of mass of the SPH
particles comes within a prescribed distance of the black hole
(effectively a tidal disruption radius). This distance is set to
$r_{tidal} = C R(M_{\rm BH}/M_{\rm NS})^{1/3}$, where {\em C} is a
constant of order unity.

\section{Results}\label{results}

We now describe our results. First, we present the initial conditions
that were used to perform the dynamical runs. We then describe the
general morphology of the coalescence events, the detailed structure
of the accretion disks that form as a result of the tidal disruption
of the neutron star, and the gravitational radiation signal.

\begin{table}
 \caption{Basic parameters for each run~(Section 3.2)}
 \label{parameters}
 \begin{tabular}{@{}ccccccc}
  Run & $q$ & $r_{RL}$ & $r_{i}$   
        & $t_{rad}$
        & $t_{f}$ & $N$ \\
  A   & 1.00 & 2.67 & 2.70 & 30.51 & 200.0 
        & 17,256 \\
  B   & 0.80 & 2.81 & 2.85 & 29.97 & 200.0 
        & 17,256 \\
  C   & 0.31 & 3.59 & 3.60 & 25.69 & 200.0 
        & 17,256 \\
  D   & 0.31 & 3.59 & 3.60 & 15.97 & 200.0 
        & 17,256 \\
  E   & 0.10 & 5.01& 5.05 & 9.56 & 200.0 
        & 17,256 \\

 \end{tabular}

 \medskip

The table lists for each run the initial mass ratio, the orbital
separation at which the Roche lobe is filled, the initial orbital
separation, the time at which gravitational radiation reaction is
switched off in the simulation, and the initial number of particles.

\end{table}

\subsection{Evolution of the Binary}\label{initial}

\begin{figure*}
\psfig{width=\textwidth,file=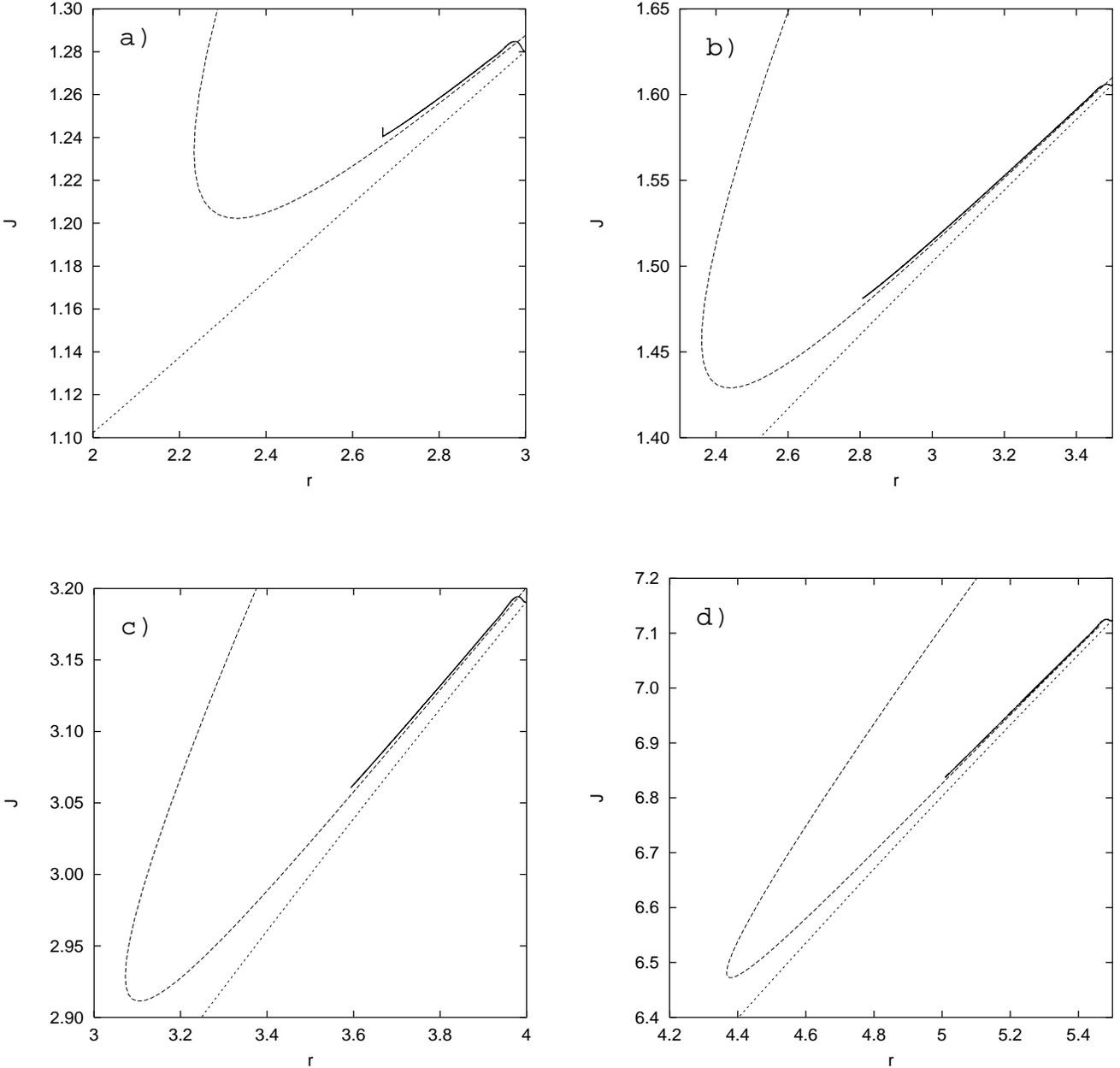,angle=0,clip=}
\caption{Total angular momentum {\em J} as a function of binary
separation {\em r} for various mass ratios. The solid line is the
result of the SPH calculation, the dashed line results from
approximating the neutron star as a compressible tri--axial ellipsoid
and the dotted line from approximating it as a rigid sphere.(a)~$q$=1;
(b)~$q$=0.8; (c)~$q$=0.31; (d)~$q$=0.1}
\label{jvsr}
\end{figure*}

To allow comparisons of results for differing equations of state, we
have run simulations with the same initial binary mass ratios as
previously explored~(Paper~I), namely $q$=1, $q$=0.8 and
$q$=0.31. Additionally we have examined the case with mass ratio
$q$=0.1. Equilibrium sequences of tidally locked binaries were
constructed for a range of initial separations, terminating at the
point where the neutron star overflows its Roche Lobe (at
$r=r_{RL}$). In Figure~\ref{jvsr}
we show the variation of total angular momentum {\em J} in these
sequences as a function of binary separation for the four values of
the mass ratio (solid lines). Following Lai, Rasio \&
Shapiro~\shortcite{LRSb}, we have also plotted the variation in {\em
J} that results from approximating the neutron star as compressible
tri--axial ellipsoid (dashed lines) and as a rigid sphere (dotted
lines). 

In all cases, the SPH results for the $\Gamma=5/3$ polytrope are very
close to the ellipsoidal approximation until the point of Roche--Lobe
overflow. This result is easy to understand if one considers that the
softer the equation of state, the more centrally condensed the neutron
star is and the less susceptible to tidal deformations arising from
the presence of the black hole. For $\Gamma=3$~(Paper~I), the
variation in angular momentum as a function of binary separation was
qualitatively different (for high mass ratios) from our present
findings. For $q$=1 and $q$=0.8, total angular momentum attained a
minimum at some critical separation {\em before } Roche--Lobe overflow
occurred. This minimum indicated the presence of a dynamical
instability, which made the binary decay on an orbital timescale. This
purely Newtonian effect arose from the tidal interactions in the
system~\cite{LRSa}. In the present study, we expect all orbits with
initial separations $r\ge r_{RL}$ to be dynamically stable.

There is a crucial difference between the two polytropes considered in
Paper~I and here.  For polytropes, the mass--radius relationship is \(
R \propto M^{(\Gamma-2)/(3\Gamma-4)}\). For $\Gamma$=3 this becomes
\(R \propto M^{1/5} \), while for $\Gamma$=5/3, \( R \propto
M^{-1/3}\).  Thus, the polytrope considered in Paper~I responded to
mass loss by shrinking. The $\Gamma=5/3$ polytrope, considered here,
responds to mass loss by expanding, as do neutron stars modeled with
realistic equations of state~\cite{arnett}--the dynamical disruption
of the star reported below seems to be related to this effect. For the
polytropic index considered in Paper~I, the star was not disrupted
(see also Lee \& Klu\'{z}niak 1995; 1997; Klu\'{z}niak \& Lee 1998),
but we find no evidence in any of our dynamical calculations for a
steady mass transfer in the binary, such as the one suggested in the
literature (e.g. Blinnikov et al.~1984; Portegies Zwart~1998).

\begin{figure*}
\psfig{width=\textwidth,file=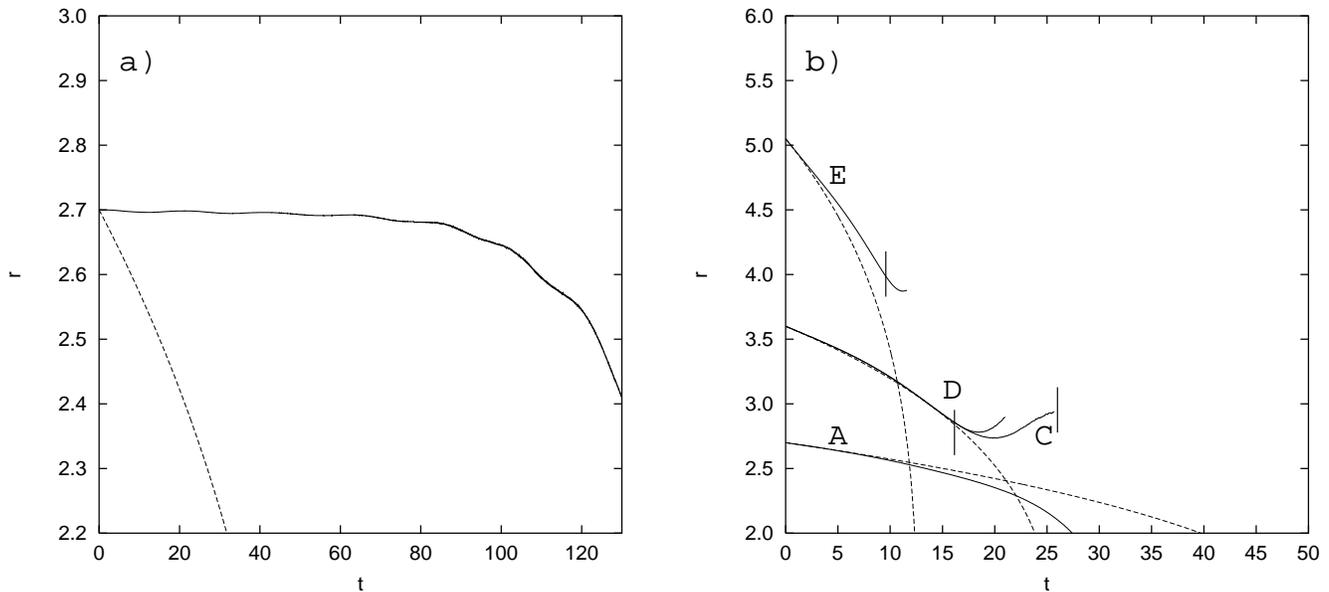,angle=0,clip=}
\caption{(a) Binary separation as a function of time for the test run
with mass ratio $q=1$ {\em without} gravitational radiation reaction
(solid line), and for a point--mass binary with the same initial
separation emitting gravitational waves, calculated in the quadrupole
approximation (dashed line).(b) Binary separation for runs A, C, D and
E (Table~\ref{parameters}) and for the corresponding point--mass
binaries. The vertical line across the solid curves marks the time
$t=t_{rad}$. The curves terminate when the axis ratio of the deformed
neutron star (in the orbital plane) is $a_{2}/a_{1} \approx 2$.}
\label{rvst}
\end{figure*}

Using equations~(\ref{eq:dedt}) and~(\ref{eq:djdt}) one can compute
the binary separation as a function of time for a point--mass binary
in the quadrupole approximation, and obtain
\begin{eqnarray}
r=r_{i}\left( 1-t/t_{0} \right)^{1/4}, \label{eq:ptdecay}
\end{eqnarray}
with \( t_{0}^{-1}=256 G^{3}M_{\rm BH} M_{\rm NS} (M_{\rm BH}+M_{\rm
NS})/(5r_{i}^{4}c^{5})\). Here $r_{i}$ is the separation at $t$=0. For
black hole--neutron star binaries studied in this paper, the timescale
for orbital decay because of angular momentum loss to gravitational
radiation, $t_{0}$, is on the order of the orbital period, $P$ (for
$q$=1, at an initial separation $r_{i}$=2.7 we find $t_{0}=56.81
\times \tilde{t}$=6.5~ms) and $P=19.58 \times \tilde{t}$=2.24~ms), so
one must analyze whether hydrodynamical effects will drive the
coalescence on a comparable timescale. We have performed a dynamical
simulation for $q$=1, with an initial separation on the verge of Roche
Lobe overflow, at $r$=2.7, {\em without} including radiation reaction
in the equations of motion. We show in Figure~\ref{rvst}a
the binary separation as a function of time for this calculation
(solid line) as well as the separation for a point--mass binary
decaying in the quadrupole approximation, using
equation~(\ref{eq:ptdecay}) (dashed line). In the dynamical simulation
the orbital separation remains approximately constant, and begins to
decay rapidly around $t$=110 (in the units defined in
equation~[\ref{eq:deftunit}]), when mass loss from the neutron star
becomes important. Clearly at this stage hydrodynamical effects are
dominant, but one must include radiation reaction in the early stages
of the process. There is an added (practical) benefit derived from
including radiation reaction in these calculations. As seen in
Figure~\ref{rvst}a, it takes a full 15~ms for the orbit to become
unstable. Simulating the behavior of the system at high resolution
(practically no SPH particles have been accreted at this stage) for
such a long time is computationally expensive, whereas accretion in
the early stages of the simulation allows us to perform, in general,
more calculations at higher resolution. We have thus included
radiation reaction in all the runs (A through E) presented in this
paper, and adopted a switch as described in section~\ref{method}.

\begin{figure*}
\psfig{width=\textwidth,file=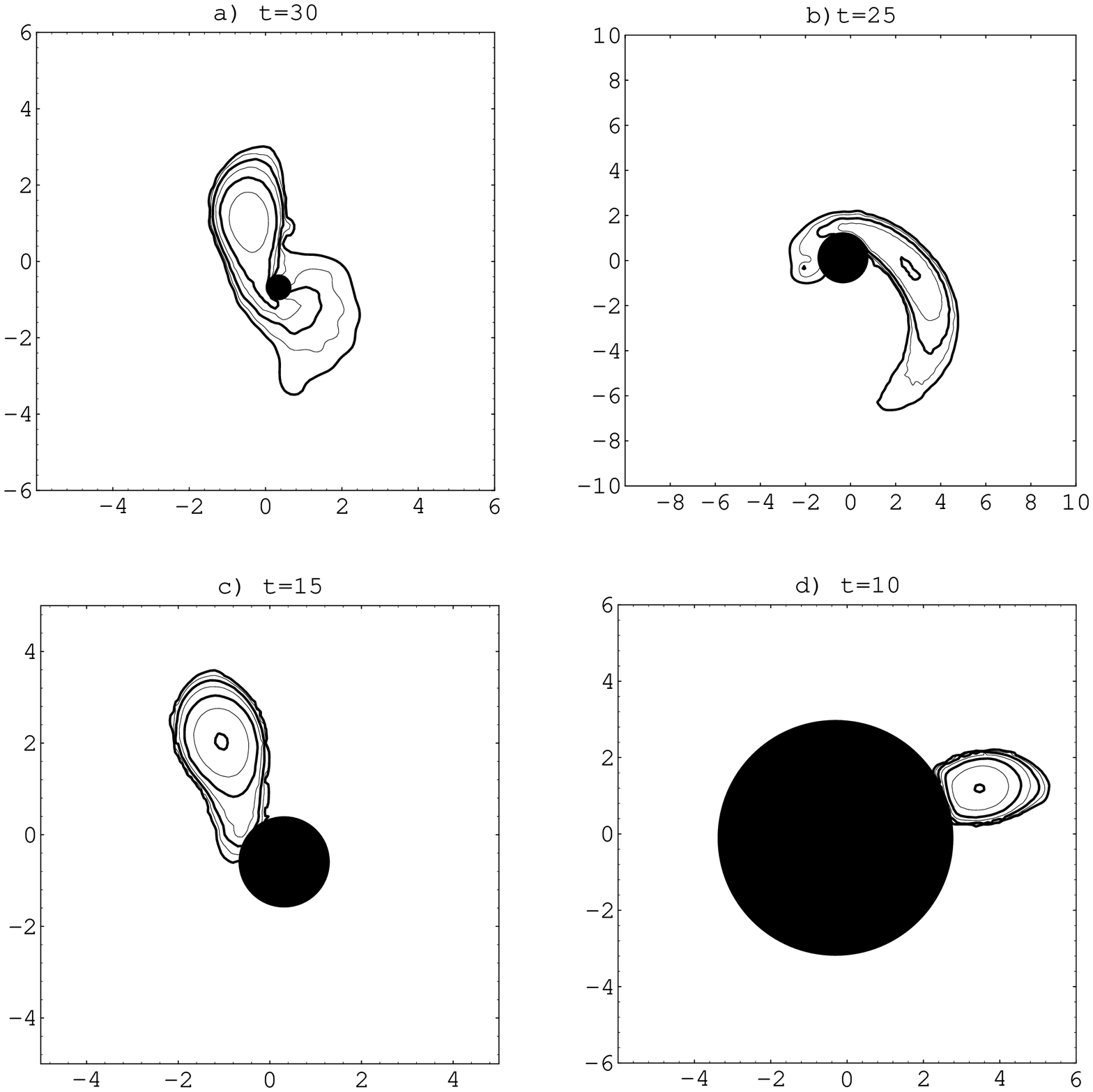,angle=0,clip=}
\caption{Density contours in the orbital plane at $t \approx t_{rad}$,
i.e., when the gravitational radiation reaction is switched off for
(a) run A, (b) run C, (c) run D, (d) run E. All contours are
logarithmic and equally spaced every 0.5 dex. The lowest (outermost)
contour is at $\log{\rho}=-3$ (in the units defined in
equation~[\ref{eq:defrhounit}]), and bold contours are plotted at
$\log{\rho}=-3,-2,-1,0$ (if present). The center of mass of the system
is at the origin of the inertial coordinate frame, all distances are
in units of the initial radius of the unperturbed star. The dark disk
of radius $r_{Sch}$ represents the black hole.}
\label{rhocontourstrad}
\end{figure*}

\begin{figure*}
\psfig{width=\textwidth,file=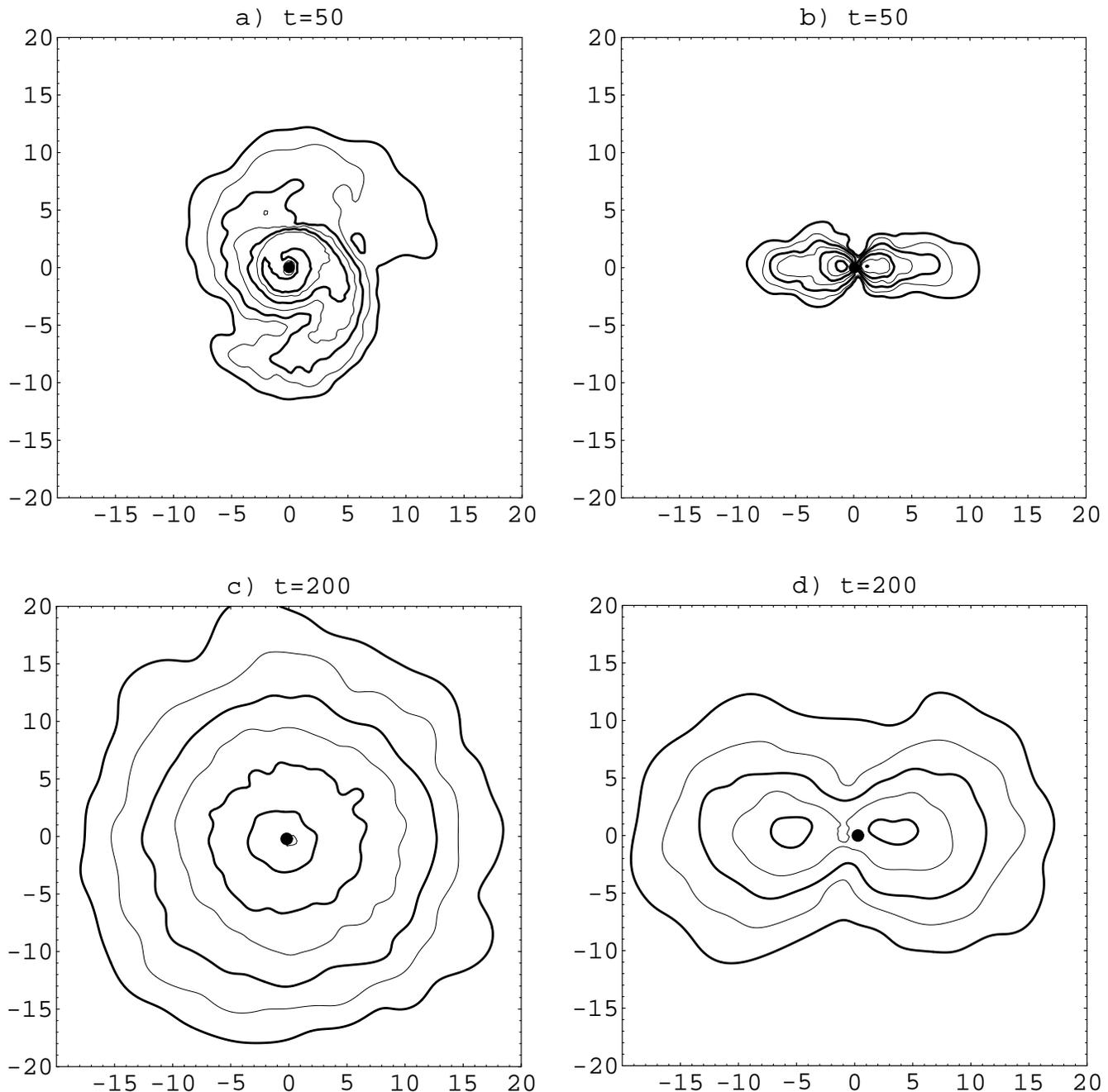,angle=0,clip=}
\caption{Density contours after the disruption of the polytrope for
$q=1$ (Run A) in the equatorial plane (left panels), and in the
meridional plane containing the black hole (right panels). All
contours are logarithmic and equally spaced every 0.5 dex. (a)--(b)
The lowest contour is at $\log{\rho}=-5$ (in the units defined in
equation~[\ref{eq:defrhounit}]), and bold contours are plotted at
$\log{\rho}=-5,-4,-3,-2$. (c)--(d) The lowest contour is at
$\log{\rho}=-6$ (in the units defined in
equation~[\ref{eq:defrhounit}]), and bold contours are plotted at
$\log{\rho}=-6,-5,-4$.}
\label{rhocontoursq1}
\end{figure*}

\begin{figure*}
\psfig{width=\textwidth,file=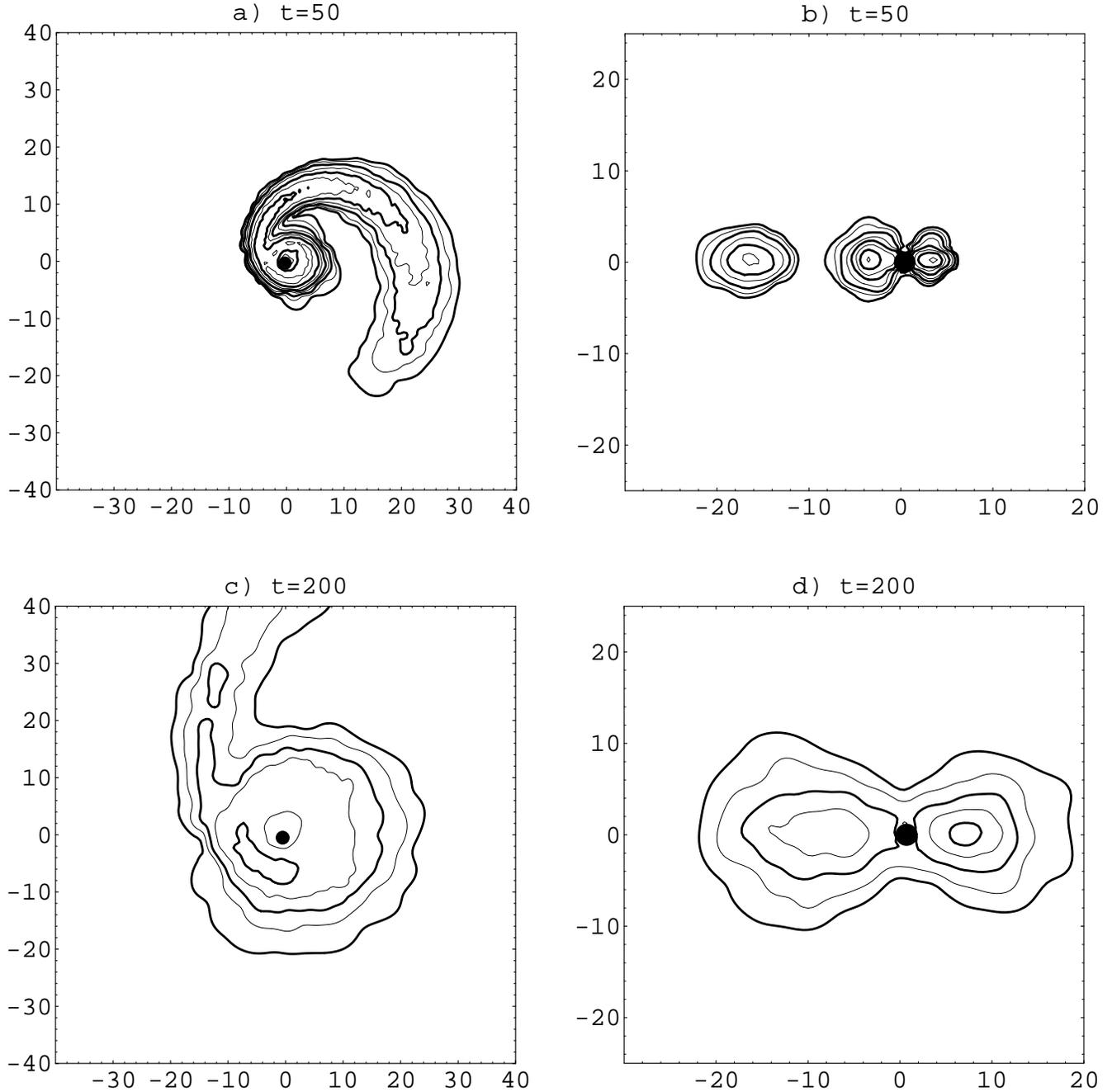,angle=0,clip=}
\caption{Same as Figure~\ref{rhocontoursq1} but for $q=0.31$ (Run
D). All contours are logarithmic and equally spaced every 0.5
dex. (a)--(d) The lowest contour is at $\log{\rho}=-6$ (in the units
defined in equation~[\ref{eq:defrhounit}]), and bold contours are
plotted at $\log{\rho}=-6,-5,-4,-3$ (if present).}
\label{rhocontoursq031b}
\end{figure*}

\begin{figure*}
\psfig{width=\textwidth,file=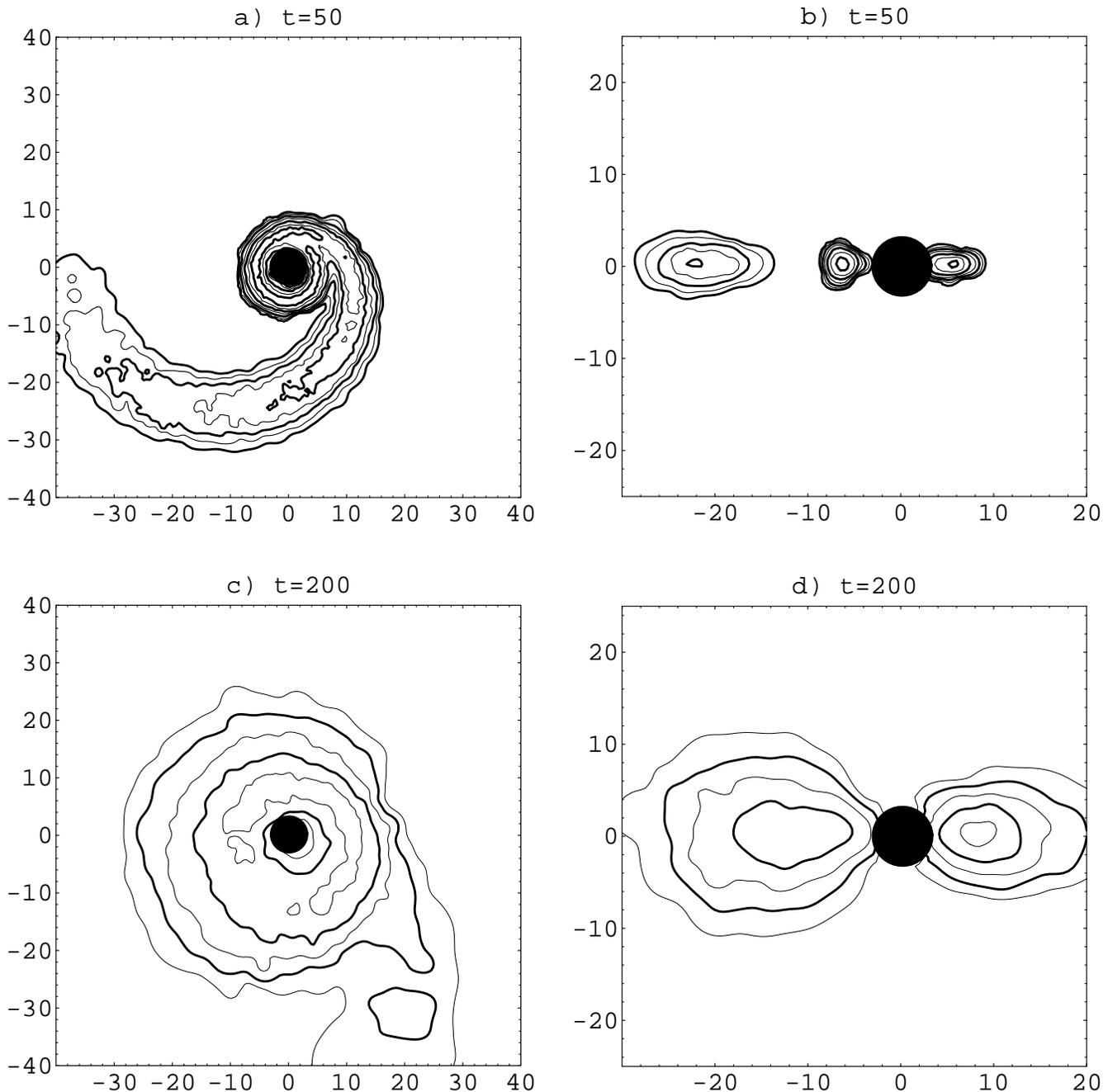,angle=0,clip=}
\caption{Same as Figure~\ref{rhocontoursq1} but for $q=0.1$ (Run
E). All contours are logarithmic and equally spaced every 0.5
dex. (a)--(b) The lowest contour is at $\log{\rho}=-6$ (in the units
defined in equation~[\ref{eq:defrhounit}]), and bold contours are
plotted at $\log{\rho}=-6,-5,-4,-3$. (c)--(d) The lowest contour is at
$\log{\rho}=-6.5$ (in the units defined in
equation~[\ref{eq:defrhounit}]), and bold contours are plotted at
$\log{\rho}=-6,-5$.}
\label{rhocontoursq01}
\end{figure*}

\begin{figure*}
\psfig{width=\textwidth,file=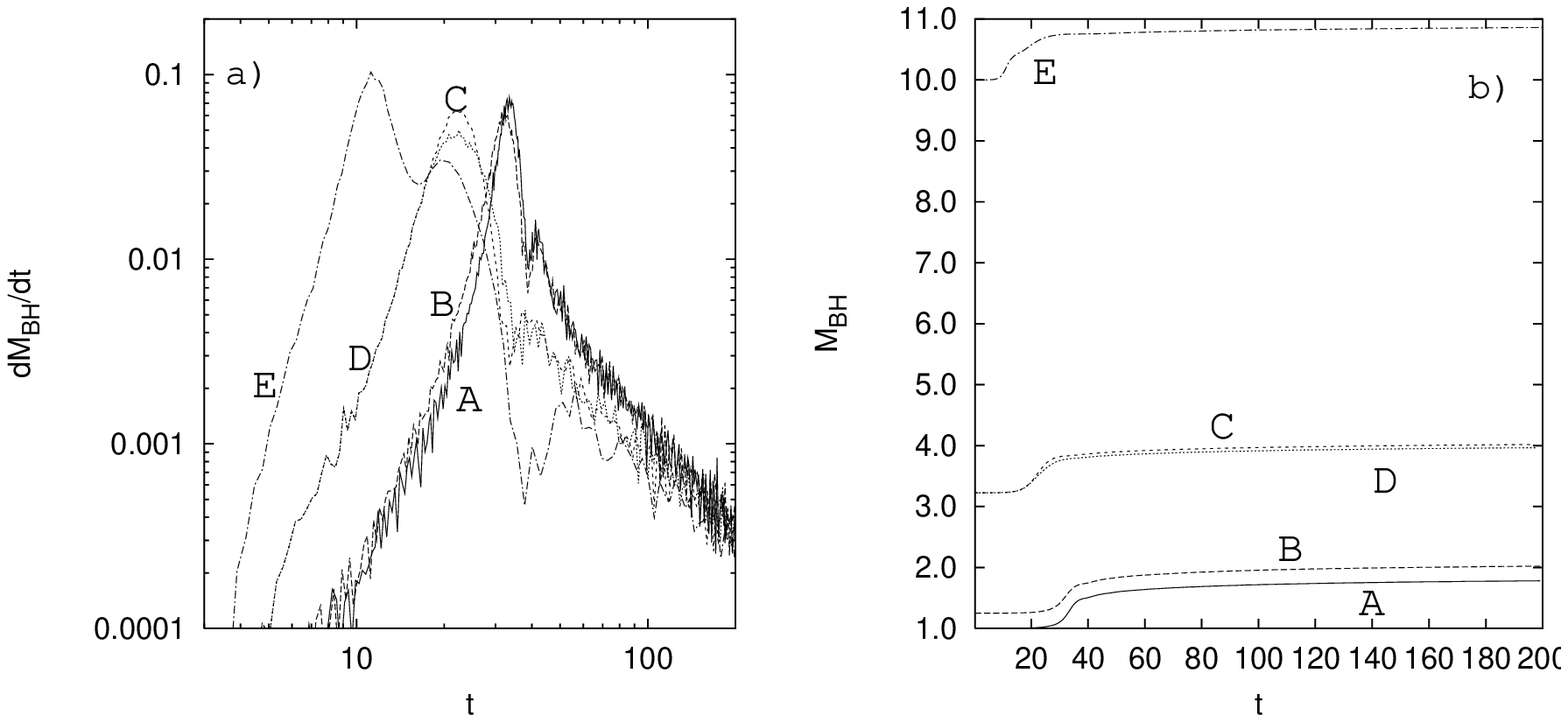,angle=0,clip=}
\caption{Mass accretion rate onto the black hole~(a) and mass of the
black hole~(b) for all runs. In both panels, the curves for runs C and
D coincide until $t$=15.97. $M_{\rm BH}$ is in units of
1.4~M$_{\odot}$ and time is in units of $\tilde{t}$ (see
equation~[\ref{eq:deftunit}]).}
\label{mdot}
\end{figure*}

\subsection{Run parameters}\label{param}

In Table~\ref{parameters} we present the parameters distinguishing
each dynamical run we performed. All times are in units of
$\tilde{t}$~(equation~[\ref{eq:deftunit}]) and all distances in units
of {\em R}, the unperturbed (spherical) stellar radius. The runs are
labeled with decreasing mass ratio (increasing black hole mass), from
$q$=1 down to $q$=0.1. All simulations were run for the same length of
time, $t_{final}=200$, equivalent to 22.9~ms (this covers on the order
of ten initial orbital periods for the mass ratios considered).

The fifth column in Table~\ref{parameters} shows the value of
$t_{rad}$, when radiation reaction is switched off according to the
criterion described in section~\ref{method}. In
Figure~\ref{rhocontourstrad} we show density contours in the orbital
plane for runs A, C, D and E at times very close to $t=t_{rad}$. The
corresponding plot for run B is very similar to that for run A. Note
that runs C and D differ {\em only} in the corresponding value of
$t_{rad}$. For run D there is little doubt that approximating the
neutron star as a point--mass is still reasonable at this stage, while
for run C this is clearly not the case. We can then use these two runs
to gauge the effect of our simple radiation reaction formulation on
the outcome of the coalescence event. We note here that run E is
probably beyond the limit of what should be inferred from a Newtonian
treatment of such a binary system. The black hole is very large
compared to the neutron star, and the initial separation
($r_{i}=5.05$, equivalent to 67.87~km) is such that the neutron star
is within the innermost stable circular orbit of a test particle
around a Schwarzschild black hole of the same mass ($r_{ms}=9.17$,
equivalent to 123.26~km). Thus we present in Appendix~\ref{pw} a
dynamical run with initial mass ratio $q=0.1$ making use of a
pseudo--Newtonian potential for the black hole. For the following, we
will at times omit a discussion of run B, as it is qualitatively and
quantitatively very similar to run A (both of these have a relatively
high mass ratio).

\subsection{Morphology of the mergers}

The initial configurations are close to Roche Lobe overflow, and mass
transfer from the neutron star onto the black hole starts within one
orbital period for all runs, A through E. Once accretion begins, the
total number of particles decreases. Since this compromises
resolution, we made a modification to the code for run E to avoid the
number of particles from dropping below 9,000. This is done simply by
splitting a given fraction of the particles $N_{split}$ and creating
$2N_{split}$ particles from them. Total mass and momentum are
conserved during this procedure, and it can be shown that the
numerical noise introduced into the smoothed density $\langle\rho
\rangle$ by doing this is of the order of the accuracy of the SPH
method itself, $\mathcal{O}$($h^{2}$), where $h$ is the smoothing
length~\cite{meglicki}.

In every run the binary separation (solid lines in in
Figure~\ref{rvst}b) initially decreases due to gravitational radiation
reaction. For high mass ratios, (runs A, B) the separation decays
faster that what would be expected of a point--mass binary. This is
also the case for a stiff equation of state, in black hole--neutron
star mergers~(Paper~I) as well as in binary neutron star
mergers~\cite{RS94}, and merely reflects the fact that hydrodynamical
effects are playing an important role. For the soft equation of state
studied here, there is the added effect of `runaway' mass transfer
because of the mass--radius relationship (see
section~\ref{initial}). For runs C and D, the solid and dashed lines
in Figure~\ref{rvst}b follow each other very closely, indicating that
the orbital decay is primarily driven by angular momentum losses to
gravitational radiation. For run E, the orbit decays more slowly than
what one would expect for a point--mass binary. This is explained by
the fact that there is a large amount of mass transfer (10\% of the
initial neutron star mass has been accreted by $t=t_{rad}$ in this
case) in the very early stages of the simulation, substantially
altering the mass ratio in the system (the dashed curves in
Figure~\ref{rvst} are computed for a fixed mass ratio). From the
expression for the timescale for orbital decay $t_{0}$ in
equation~(\ref{eq:ptdecay}), it is apparent that at constant total
mass, lowering the mass ratio slows the orbital decay, when $q<0.5$.

The general behavior of the system is qualitatively similar for every
run. Figures~\ref{rhocontoursq1},~\ref{rhocontoursq031b} and
~\ref{rhocontoursq01} show density contours in the orbital plane (left
columns) and in the meridional plane containing the black hole (right
columns) for runs A, D and E respectively, at $t=50$ and $t=t_{f}=200$
(equivalent to 5.73~ms and 22.9~ms). The corresponding plots for runs
B and C are very similar to those for runs A and D, respectively. The
neutron star becomes initially elongated along the binary axis and an
accretion stream forms, transferring mass to the black hole through
the inner Lagrange point. The neutron star responds to mass loss and
tidal forces by expanding, and is tidally disrupted. An accretion
torus forms around the black hole as the initial accretion stream
winds around it. A long tidal tail is formed as the material furthest
from the black hole is stripped from the star. Most of the mass
transfer occurs in the first two orbital periods and peak accretion
rates reach values between 0.04 and 0.1---equivalent to 0.49 and 1.22
M$_{\odot}$/ms (see Figure~\ref{mdot}). The mass accretion rate then
drops and the disk becomes more and more azimuthally symmetric,
reaching a quasi--steady state by the end of the simulations.

\begin{figure*}
\psfig{width=\textwidth,file=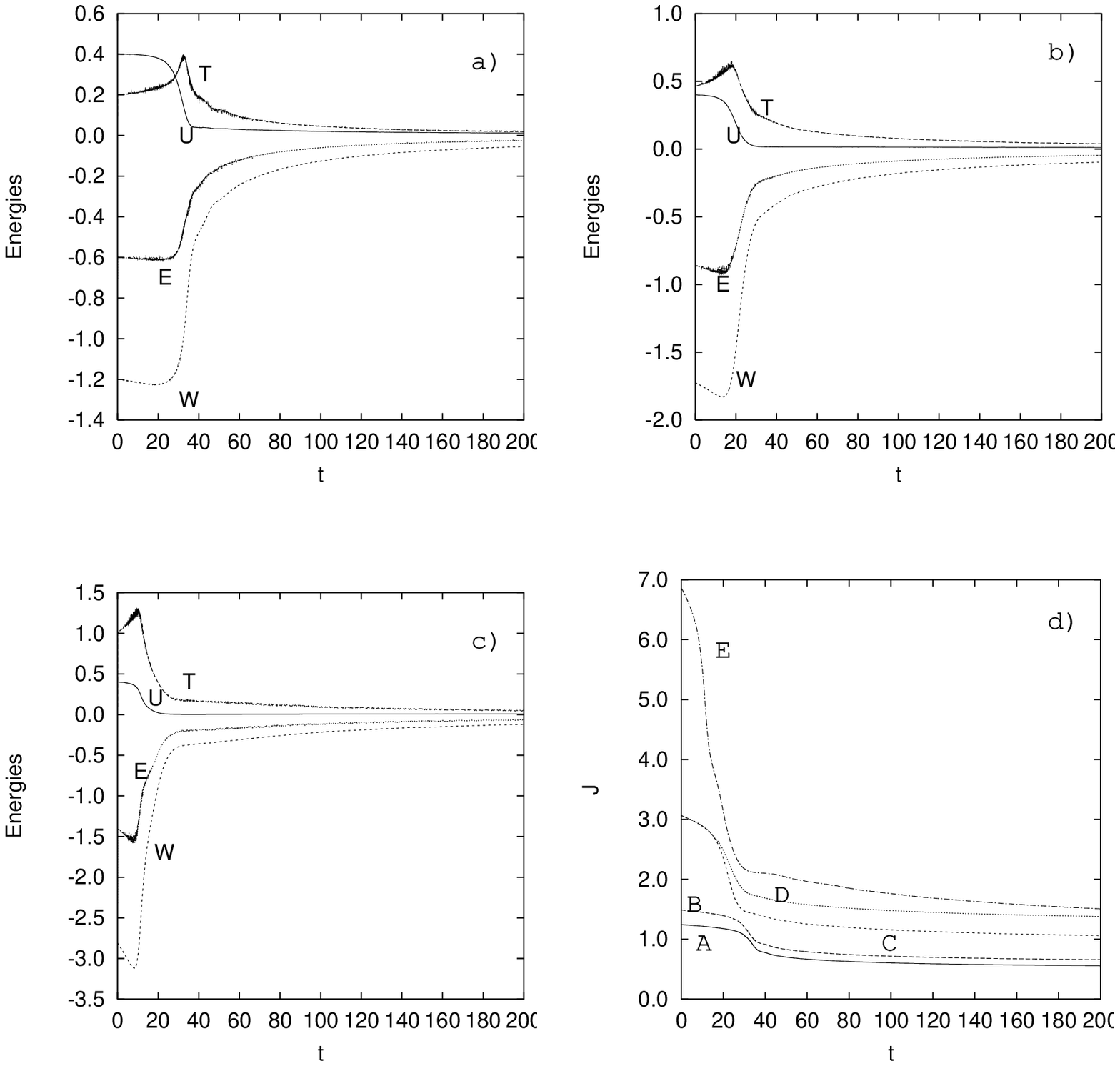,angle=0,clip=}
\caption{(a) Various energies of the system as a function of time for
run A, (b) energies in run D, (c) energies in run E. The kinetic (T),
internal (U), gravitational potential (W) and total (E) energies are
indicated in units of $3.8\times 10^{53}$~erg. (d) Angular momentum as
a function of time for every run in units of $4.37\times
10^{42}$~kg~m$^{2}$~s$^{-1}$.}
\label{energies}
\end{figure*}

We show in Figure~\ref{energies} the various energies of the system
(kinetic, internal, gravitational potential and total) for runs A, D
and E. 
The dramatic drop in total internal energy reflects the intense mass
accretion that takes place within the first couple of
orbits. Figure~\ref{energies} also shows [in panel (d)] the total
angular momentum of the system (the only contribution to the total
angular momentum not plotted is the spin angular momentum of the black
hole, see below). Angular momentum decreases for two reasons. First,
if gravitational radiation reaction is still acting on the system, it
will decrease approximately according to
equation~(\ref{eq:djdt}). Second, whenever matter is accreted by the
black hole, the corresponding angular momentum is removed from our
system. In reality, the angular momentum of the accreted fluid would
increase the spin of the black hole. We keep track of this accreted
angular momentum and exhibit its value in Table~2 as the Kerr
parameter of the black hole. This shows up as a decrease in the total
value of {\em J}. It is clear from runs C and D that the value of
$t_{rad}$ influences the peak accretion rate and the mass of the black
hole (particularly immediately after the first episode of heavy mass
transfer). The maximum accretion rate is different for run C and D by
about a factor of 1.4. This is easy to understand, since radiation
reaction increases angular momentum losses, and hence more matter is
accreted per unit time when it is present.

\subsection{Accretion disk structure}

\begin{figure}
\psfig{width=85mm,file=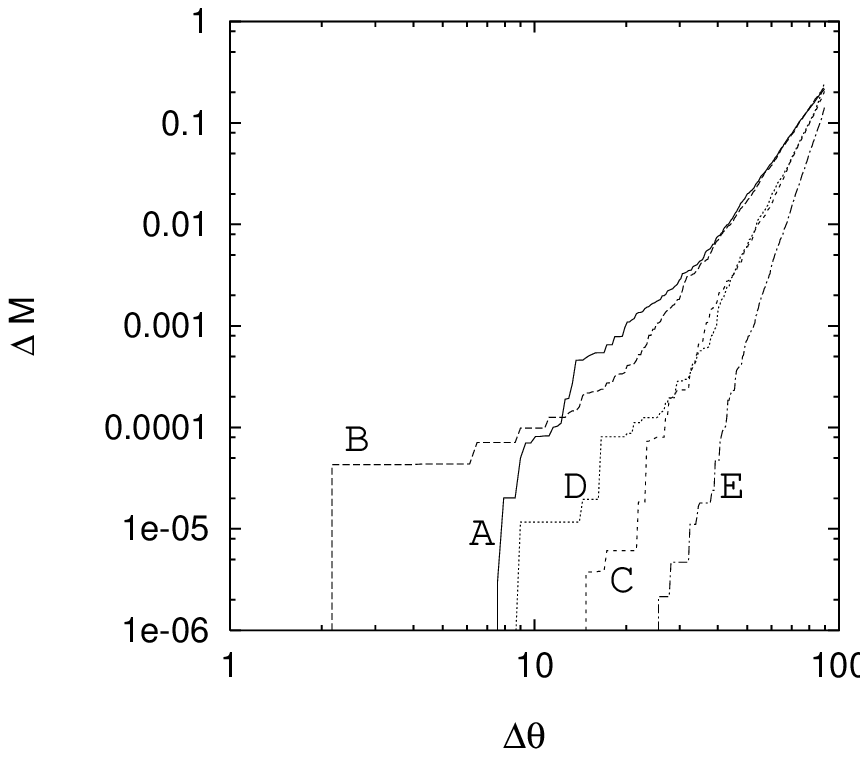,angle=0,clip=}
\caption{Enclosed mass as a function of half--angle $\Delta \theta$
(measured from the rotation axis in degrees) for all runs at
$t=t_{f}$. The mass resolution varies from approximately $5 \times
10^{-6}$ to $5 \times 10^{-5}$ (corresponding to $7 \times
10^{-6}$~M$_{\odot}$ and $7 \times 10^{-5}$~M$_{\odot}$ respectively)
in the region directly above the black hole.}
\label{mtheta}
\end{figure}

In Table~\ref{disks} we show several parameters pertaining to the
final accretion structure around the black hole for every run. The
disk settles down to a fairly azimuthally symmetric structure within a
few initial orbital periods (except for the long tidal tail, which
always persists as a well--defined structure), and there is a
baryon--free axis above (and below) the black hole in every case
(Figure~\ref{mtheta}). We have calculated the mass of the remnant
disk, $M_{disk}$, by searching for the amount of matter that has
sufficient specific angular momentum {\em j} at the end of the
simulation to remain in orbit around the black hole (as in Ruffert \&
Janka~1999). This material has $j>j_{crit}=\sqrt{6}GM_{t}/c$, where
$M_{t}$ is the total mass of the system. The values shown in
Table~\ref{disks} are equivalent to a few tenths of a solar mass, and
again the effect of $t_{rad}$ can be seen by comparing runs C and D,
where the disk masses differ by a factor of 1.2. By the end of the
simulations, between 70\% and 80\% of the neutron star has been
accreted by the black hole.  It is interesting to note that the {\em
final} accretion rate (at $t=t_{f}$) appears to be rather insensitive
to the initial mass ratio, and is between $2\times 10^{-4}$ and
$5\times 10^{-4}$ (equivalent to 2.4 and 6.1 M$_{\odot}$~s$^{-1}$
respectively). From this final accretion rate we have estimated a
typical timescale for the evolution of the accretion disk,
$\tau_{disk}=M_{disk}/\dot{M}_{final}$---for reference, $\tau=100$, in
the units of equation~(\ref{eq:deftunit}) corresponds to $11.5$~ms and
thus the values of $\tau_{disk}$ given in Table~\ref{disks} are
between 47 and 63~ms. Despite the difference in the initial mass
ratios and the typical sizes of the disks ($r_{0}$ is the radial
distance from the black hole to the density maximum at $t=t_{f}$), the
similar disk masses and final accretion rates make the lifetimes
comparable for every run.

\begin{figure*}
\psfig{width=\textwidth,file=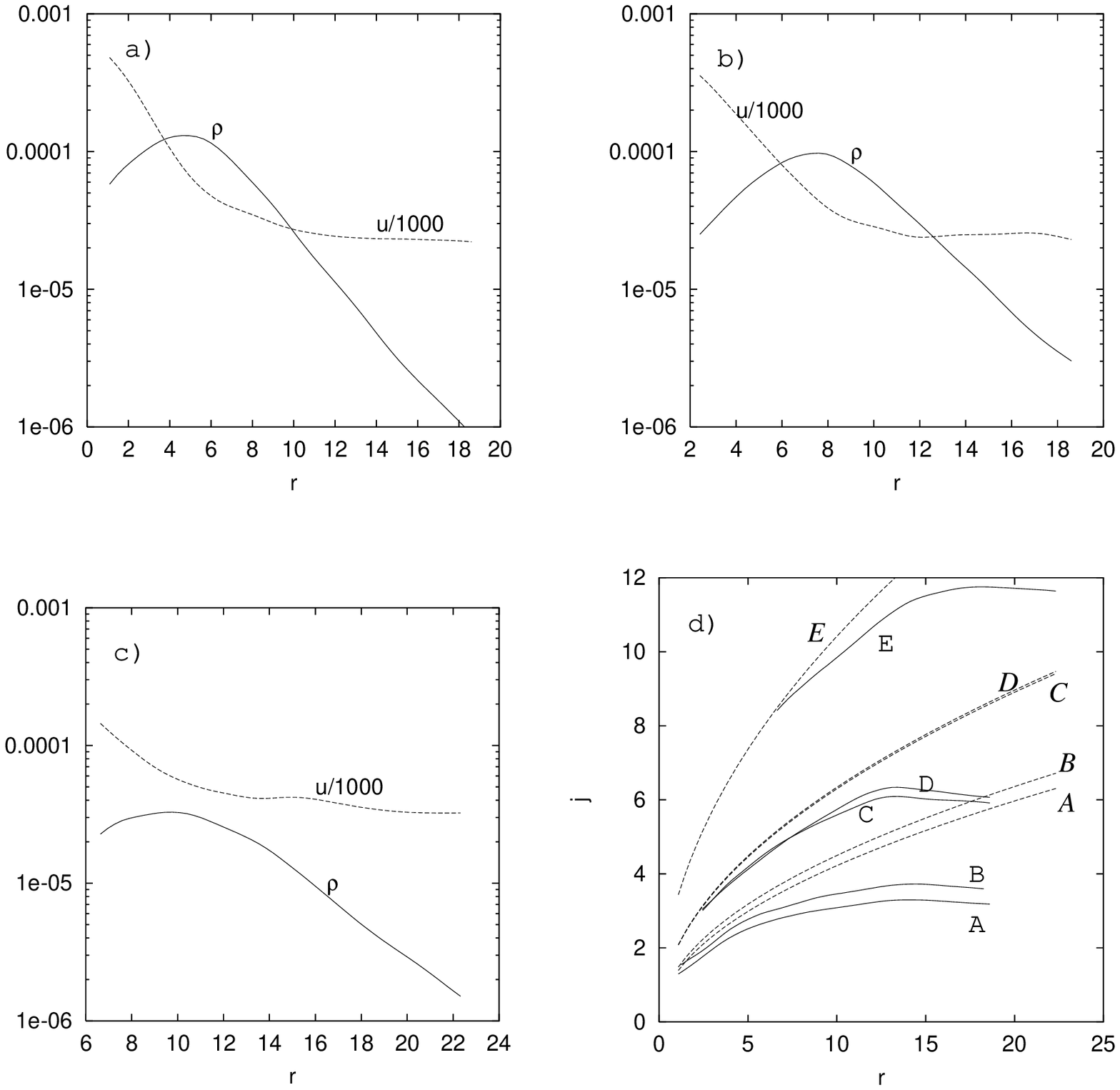,angle=0,clip=}
\caption{Azimuthally averaged profiles for the density, $\rho$, and
the specific internal energy, {\em u} ($u/1000$ is plotted), of the
accretion torus at $t=t_{f}$ for runs A~(a), D~(b) and E~(c). The
inner edge of the curves is at $r=2 r_{Sch}$. At this stage in the
simulation the tori are close to being azimuthally symmetric. For
reference, $\rho=10^{-4}$ corresponds to $1.14 \times
10^{11}~\mbox{g~cm$^{-3}$}$ and $u/1000=10^{-4}$ corresponds to $1.37
\times 10^{19}~\mbox{erg~g$^{-1}$}$. Panel~(d) shows the distribution
of specific angular momentum {\em j} for all runs at $t=t_{f}$ (solid
lines, A, B, C, D, E) and the specific angular momentum of a Keplerian
accretion disk for the same black hole mass (dashed lines, {\it A},
{\it B}, {\it C}, {\it D}, {\it E}).}
\label{diskprofiles}
\end{figure*}

We have plotted azimuthally averaged density and internal energy
profiles in Figure~\ref{diskprofiles} for runs A, D and E. The
specific internal energy is greater towards the center of the disk,
and flattens out at a distance from the black hole roughly
corresponding the density maximum, at $u\simeq 2 \times 10^{-2}$. This
value corresponds to $2.74 \times 10^{18}$~erg~g$^{-1}$ or
2.9~MeV/nucleon and is largely independent of the initial mass
ratio. The inner regions of the disks have specific internal energies
that are greater by approximately one order of magnitude.

Additionally, panel (d) in the same figure shows the azimuthally
averaged distribution of specific angular momentum {\em j} in the
orbital plane for all runs. The curves terminate at $r_{in}=2r_{Sch}$.
Pressure support in the inner regions of the accretion disks makes the
rotation curves sub--Keplerian, while the flattening of distribution
marks the outer edge of the disk and the presence of the long tidal
tail (see Figure~\ref{ejected}),which has practically constant
specific angular momentum.

The Kerr parameter of the black hole, given by $a=J_{\rm BH}c/G M_{\rm
BH}^{2}$, is also shown in Table~\ref{disks}. We have calculated it
from the amount of angular momentum lost via accretion onto the black
hole (see Figure~\ref{energies}d), assuming that the black hole is
{\em not} rotating at $t=0$. The final specific angular momentum of
the black hole is smaller for lower mass ratios simply because the
black hole is initially more massive when {\em q} is smaller. The
difference in the value of {\em a} for runs C and D is important
(almost a factor of 2), and again reflects the influence of
gravitational radiation reaction (for a larger value of $t_{rad}$ the
black hole is spun up to a greater degree because of the larger amount
of accreted mass).

It is of crucial importance for the production of GRBs from such a
coalescence event that there be a baryon--free axis in the system
along which a fireball may expand with ultrarelativistic
velocities~(M\'{e}sz\'{a}ros \& Rees 1992; 1993). We have calculated
the baryon contamination for every run as a function of the
half--angle $\Delta \theta$ of a cone directly above (and below) the
black hole and along the rotation axis of the binary that contains a
given amount of mass $\Delta M$. Table~\ref{disks} shows these angles
(in degrees) for $\Delta M=10^{-3},10^{-4},10^{-5}$ (equivalent to
$1.4\times 10^{-3},1.4\times 10^{-4},1.4\times 10^{-5}$~M$_{\odot}$
respectively). There is a greater amount of pollution for high mass
ratios (the disk is geometrically thicker compared to the size of the
black hole), but in all cases only modest angles of collimation are
required to avoid contamination. We note here that the values for
$\theta_{-5}$ are rough estimates at this stage since they are at the
limit of our numerical resolution in the region directly above the
black hole. This can be seen by inspection in Figure~\ref{mtheta}
where we show the enclosed mass as a function of half--angle $\Delta
\theta$ for all runs at $t=t_{f}$.

\begin{table*}
 \caption{Accretion disk structure}
 \label{disks}
 \begin{tabular}{@{}cccccrrcccccc}
  Run & $q$ & $M_{disk}$ & $M_{acc}$ & $\dot{M}_{max}$    
	& $\dot{M}_{final}$ 
        & $M_{ejected}$ 
        & $r_{0}$ & $\tau_{disk}$ & $J_{\rm BH}c/G M_{\rm BH}^{2}$ &
        $\theta_{-3}$
        & $\theta_{-4}$ & $\theta_{-5}$ \\
  A   & 1.00 & 0.188 & 0.78 & 0.068 & $4 \cdot 10^{-4}$ & $9.51 \cdot 10^{-3}$ 
& 4.83 & 472 
        & 0.517 & 20 & 12 & 8 \\
  B   & 0.80 & 0.198 & 0.77 & 0.060 & $5 \cdot 10^{-4}$ & $5.41 \cdot 10^{-3}$ 
& 4.46 & 409 
        & 0.497 & 25 & 10 & 3 \\
  C   & 0.31 & 0.184 & 0.79 & 0.062 & $3.5 \cdot 10^{-4}$ & $3.95 \cdot 10^{-3}$
  & 6.32 & 532 
        & 0.313 & 37 & 27 & 22 \\
  D   & 0.31 & 0.226 & 0.74 & 0.045 & $4 \cdot 10^{-4}$ & $16.12 \cdot 10^{-3}$
 & 7.44 & 551 
        & 0.173 & 40 & 21 & 10 \\
  E   & 0.10 & 0.072 & 0.86 & 0.095 & $2 \cdot 10^{-4}$ & $3.18 \cdot 10^{-3}$ 
& 10.41 & 410 
        & 0.114 & 52 & 42 & 32 \\

 \end{tabular}

 \medskip

In the last three columns, $\theta_{-n}$ is the half--angle of a cone
above the black hole and along the rotation axis of the binary that
contains a mass $M=10^{-n}$.

\end{table*}

\subsection{Ejected mass and r--process}

\begin{figure*}
\psfig{width=\textwidth,file=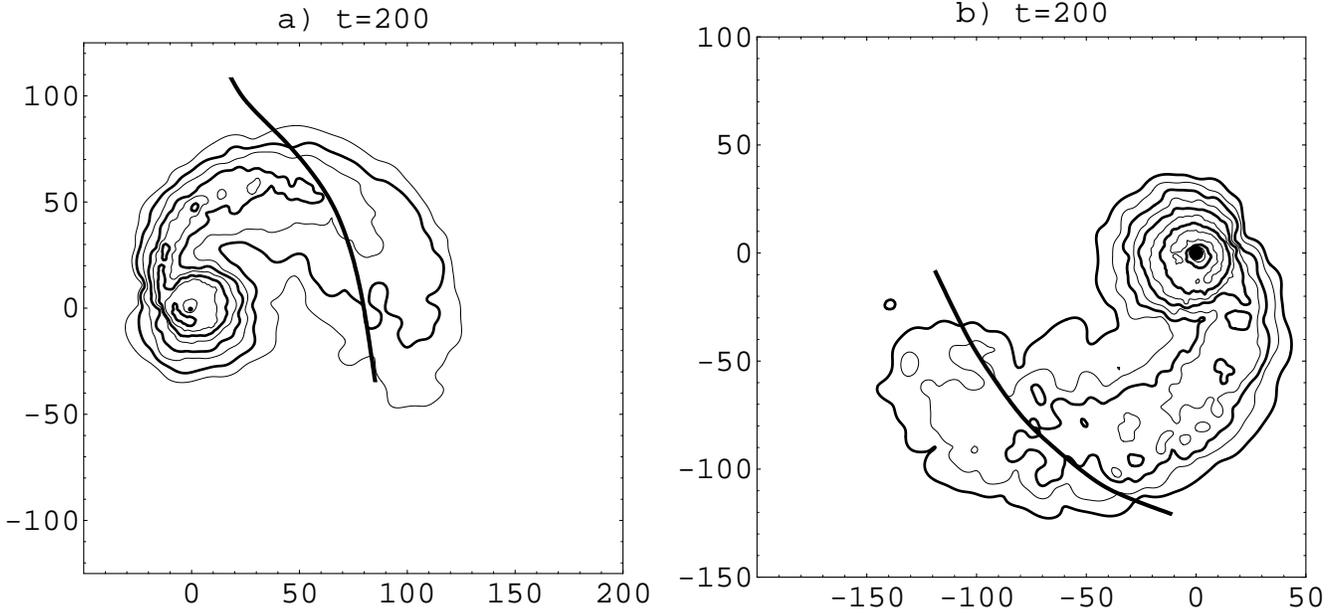,angle=0,clip=}
\caption{Final density contours in the orbital plane for runs D (left
panel) and E (right panel). All contours are logarithmic and equally
spaced every 0.5 dex. (a) The lowest contour is at $\log{\rho}=-7.5$
(in the units defined in equation~[\ref{eq:defrhounit}]), and bold
contours are plotted at $\log{\rho}=-7,-6,-5,-4$. (b) The lowest
contour is at $\log{\rho}=-8$ (in the units defined in
equation~[\ref{eq:defrhounit}]), and bold contours are plotted at
$\log{\rho}=-8,-7,-6,-5$. The thick black line running across the
tidal tail in each frame divides the matter that is bound to the black
hole from that which is on outbound trajectories.}
\label{ejected}
\end{figure*}

To calculate the amount of dynamically ejected mass during the
coalescence process, we look for matter that has a positive total
energy (kinetic+gravitational potential+internal) at the end of each
simulation. Figure~\ref{ejected} shows a large--scale view of the
system at $t=t_{f}$ for runs D and E. The thick black line running
across the tidal tail in each case divides matter that is bound to the
black hole from that which may be on outbound trajectories. This
matter comes from the part of the neutron star that was initially
furthest from the black hole and was ejected through the outer
Lagrange point in the very early stages of mass transfer. We find that
a mass between $4.4 \times 10^{-3}$~M$_{\odot}$ and $2.2 \times
10^{-2}$~M$_{\odot}$ can potentially be ejected in this fashion (see
Table~\ref{disks}). This is very similar to what has recently been
calculated for binary neutron star mergers for a variety of initial
configurations~\cite{rosswog}. Since it is believed that the event
rate for binary neutron star mergers is comparable to that of black
hole--neutron star mergers, this could prove to be a sizable
contribution to the amount of observed r--process material. This
result appears to be strongly dependent on the equation of state,
since we previously observed no significant amount of matter being
ejected for a system with a stiff equation of state~(Paper~I). For
binary neutron star mergers, Rosswog et al.~1999 have obtained the
same qualitative result. We also note that there is a significant
difference in the amount of ejected mass for runs C and D
(approximately a factor of four), due to the difference in the values
of $t_{rad}$ (in run D the system loses less angular momentum and thus
matter escapes with greater ease).

\subsection{Gravitational radiation waveforms and luminosities}

\begin{figure*}
\psfig{width=\textwidth,file=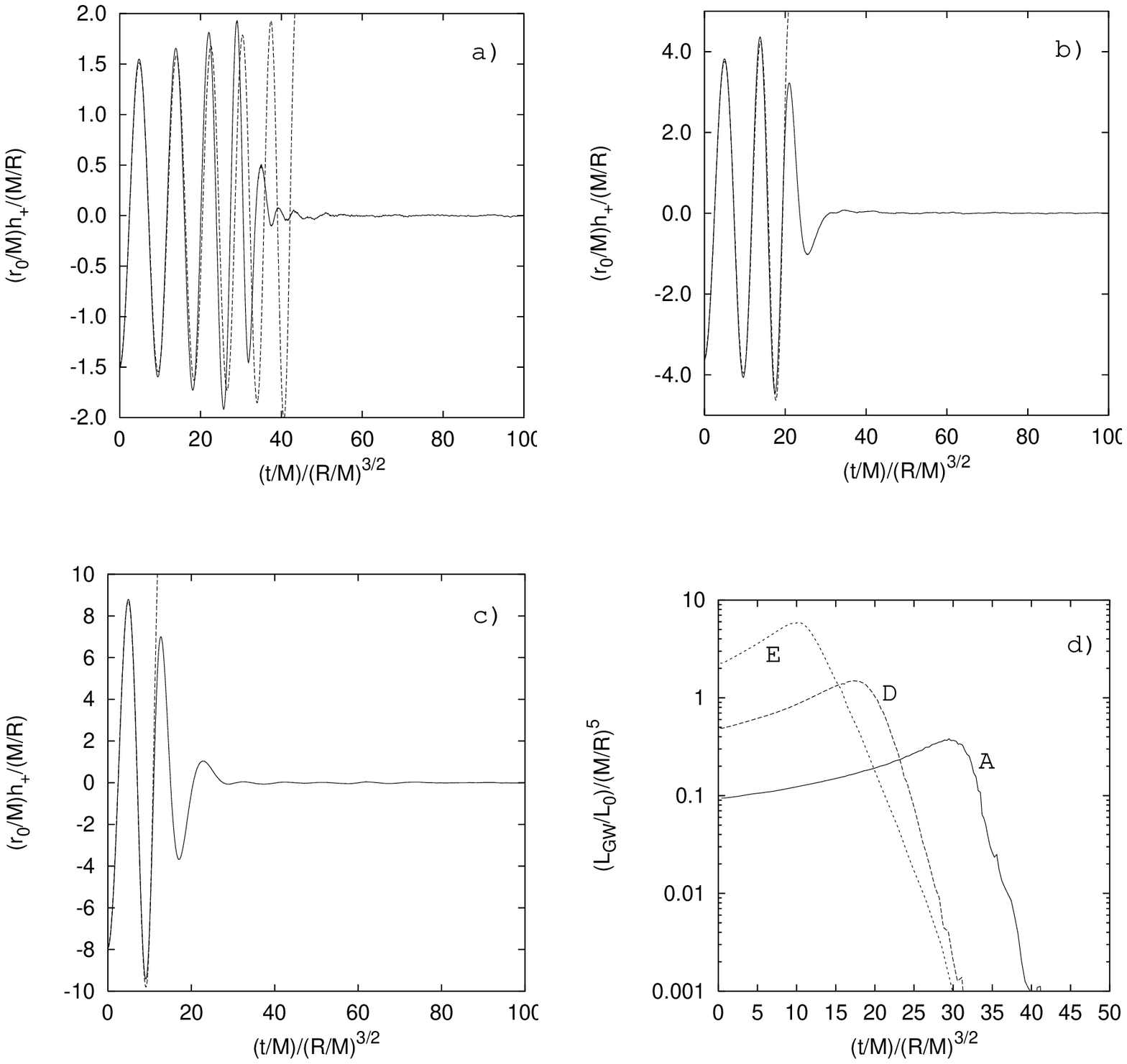,angle=0,clip=}
\caption{(a)--(c):Gravitational radiation waveforms (one polarization)
for runs A~(a), D~(b) and E~(c). The solid lines are the results from
the dynamical simulations, while the dashed lines show the emission
from a point--mass binary with the same initial separation and mass
ratio, calculated in the quadrupole approximation.(d) Gravitational
radiation luminosity. All quantities are plotted in geometrized units
such that $G$=$c$=1 ($L_{0}=c^{5}/G=3.64 \times
10^{59}\mbox{erg~s$^{-1}$}$).}
\label{gw}
\end{figure*}

\begin{table*}
 \caption{Gravitational radiation}
 \label{waves}
 \begin{tabular}{@{}ccccc}
  Run & $q$ & $(r_{0}R/M_{\rm NS}^{2})h_{max}$   
        & $(R/M_{\rm NS})^{5}(L_{max}/L_{0})$
        & $(R^{7/2}/M_{\rm NS}^{9/2})\Delta E_{GW}$ \\
      &      &      &      &      \\
  A   & 1.00 & 1.93 & 0.37 & 6.53 \\ 
  B   & 0.80 & 2.26 & 0.44 & 8.24 \\
  C   & 0.31 & 4.40 & 1.45 & 21.11 \\
  D   & 0.31 & 4.46 & 1.50 & 20.77 \\
  E   & 0.10 & 9.42 & 5.90 & 60.08 \\

 \end{tabular}

 \medskip 

All quantitites are given in geometrized units such that $G=c=1$, and
$L_{0}=c^{5}/G=3.64\times 10^{59}$~erg~s$^{-1}$.
\end{table*}

The emission of gravitational radiation is calculated in all our
models in the quadrupole approximation (see e.g. Finn~1989; Rasio \&
Shapiro~1992), and can be obtained directly from the hydrodynamical
variables of the system. The calculation of the gravitational
radiation luminosity then requires only an additional numerical
differentiation. Figure~\ref{gw} shows the computed waveforms and
luminosities, along with what the waveforms would be for a point--mass
binary, also calculated in the quadrupole approximation. It is
apparent that hydrodynamical effects play an important role
particularly for high mass ratios in the early stages of the
coalescence process (see panel~(a) in Figure~\ref{gw}). When the
neutron star is tidally disrupted, the amplitude of the waveform drops
abruptly and practically to zero, since a structure that is almost
azimuthally symmetric has formed around the black hole. This is in
stark contrast to what occurred for a stiff equation of
state~(Lee~1998; Paper~I; Klu\'{z}niak \& Lee~1998), when the binary
system survived the initial episode of mass transfer and a stable
binary was the final outcome. In fact, these waveforms resemble more
the case of a double neutron star merger with a soft equation of
state~\cite{RS92}, in which the coalescence resulted in a compact,
azimuthally symmetric object surrounded by a dense halo and spiral
arms. Table~\ref{waves} shows the maximum amplitude $h_{max}$ for an
observer located a distance $r_{0}$ away from the system along the
axis of rotation, the maximum luminosity $L_{max}$ and the total enery
$\Delta E_{GW}$ emitted during the event. This last number should be
taken only as an order of magnitude estimate since it depends on the
choice of the origin of time. The peak luminosities are $(R/M_{\rm
NS})^{5}(L_{max}/L_{0})=0.37$ for run A, $(R/M_{\rm
NS})^{5}(L_{max}/L_{0})=1.50$ for run D and $(R/M_{\rm
NS})^{5}(L_{max}/L_{0})=5.90$ for run E (equivalent to $ 1.12 \times
10^{55}$erg~s$^{-1}$, $4.55 \times 10^{55}$erg~s$^{-1}$ and $ 1.79
\times 10^{56}$erg~s$^{-1}$ respectively). We note that although the
waveforms for runs C (not plotted) and D (panel (b) in
Figure~\ref{gw}) are very similar, the maximum amplitudes and the
luminosity differ by about 1.3\% and 3.4\% respectively, a small but
non--negligible amount. This is again a reflection of the way in which
the radiation reaction was formulated, and indicates that a more
rigorous treatment of this effect is necessary.

\section{Summary and Discussion}\label{discussion}

We have presented results of hydrodynamical simulations of the binary
coalescence of a black hole with a neutron star. We have used a
polytropic equation of state (with index $\Gamma=5/3$) to model the
neutron star, and a Newtonian point mass with an absorbing surface at
the Schwarzschild radius to represent the black hole. All our
computations are strictly Newtonian, but we have included a term that
approximates the effect of gravitational radiation reaction in the
system. We have also calculated the emission of gravitational
radiation in the quadrupole approximation.

We have found that for every mass ratio investigated ($M_{\rm
NS}/M_{\rm BH}$=1, 0.8, 0.31 and 0.1) the $\Gamma=5/3$ polytrope
(`neutron star') is entirely disrupted by tidal forces, and a dense
accretion torus, containing a few tenths of a solar mass, forms around
the black hole. The maximum densities and specific internal energies
in the tori are on the order of $10^{11}~\mbox{g~cm$^{-3}$}$ and
$10^{19}~\mbox{erg~g$^{-1}$}$ (or 10~MeV/nucleon) respectively (all
simulations were run for approximately 22.9~ms). The final accretion
rate is between 2 and 6 solar masses per second, and hence the
expected lifetime of the torus $\tau_{disk}=M_{disk}/\dot{M}_{final}$
is between 40 and 60 milliseconds.

The rotation axis of the system remains free of matter to a degree
that would not hinder the production of a relativistic fireball,
possibly giving rise to a gamma ray burst. Although the duration of
such a bursts would still be too short to power the longest GRBs, the
present scenario could well account for the subclass of short
bursts~\cite{kouveliotou}. A significant amount of matter (between
$10^{-2}$ and $10^{-3}$ solar masses) is dynamically ejected from the
system, and could contribute significantly to the observed abundances
of r--process material in our galaxy. The gravitational radiation
signal is very similar to that of a point--mass binary until the
beginning of mass transfer, particularly for low mass ratios. After
mass transfer starts, the amplitude of the waveforms drops
dramatically on a dynamical timescale when the accretion torus is
formed. In every aspect, the results are dramatically different from
what occurs for a stiff equation of state.

\section*{Acknowledgments}

We gratefully acknowledge financial support from DGAPA--UNAM and KBN
grant 2P03D01311. W.L. thanks Craig Markwardt for helpful discussions
concerning the effect of a soft equation of state on the system. It is
a pleasure to thank the referee for his most helpful comments.

\appendix

\section{Dynamical evolution for a pseudo--Newtonian potential}\label{pw}

Since, as stated in section~\ref{param}, the system with mass ratio
$q=0.1$ (run E) is at an initial separation that is within the
marginally stable orbit for test particles around a Schwarzschild
black hole, we have performed an additional run, altering the form of
the potential produced by the black hole. We will perform a detailed
description of results elsewhere (Lee~1999). Here we only present a
comparison to the results of run E. We have chosen the form proposed
by Paczy\'{n}ski \& Wiita~(1980), namely:
\begin{eqnarray}
\Phi^{PW}_{\rm BH}(r) = -GM_{\rm BH}/(r-r_{Sch}).
\end{eqnarray}
This potential correctly reproduces the positions of the marginally
bound and marginally stable orbits for test particles. A few
modifications need to be made to the SPH code to accomodate this new
potential. First, the absorbing boundary of the black hole is now
placed at a distance $r_{boundary}=1.5 r_{Sch}$ from the position of
the black hole; second, the total gravitational force exerted by the
neutron star on the black hole is symmetrized so that total linear
momentum is conserved in the system. 

A tidally locked equilibrium configuration for a given separation $r$
can be constructed with these modifications in the same manner as
described in section~\ref{initial}. For a test particle in orbit about
a Schwarzschild black hole, the marginally stable orbit appears at a
separation $r_{ms}=6GM_{\rm BH}/c^{2}$ because the total angular
momentum exhibits a minimum at that point. In our case (where the
neutron star has a finite size and mass), the turning point in the
curve of total angular momentum as a function of binary separation
occurs at approximately $r=9.1 R_{\rm NS}$, and so we have chosen this
value for the initial separation $r_{i}$ to be used in the dynamical
simulation. Gravitational radiation reaction has been implemented as
described in section~\ref{method}, with a slight modification to the
definition of $r_{tidal}$ to account for the increased strength of the
gravitational interactions, so that now $r_{tidal}=CR(M_{\rm
BH}/M_{\rm NS})^{1/3}+r_{Sch}$.

\begin{figure*}
\psfig{width=\textwidth,file=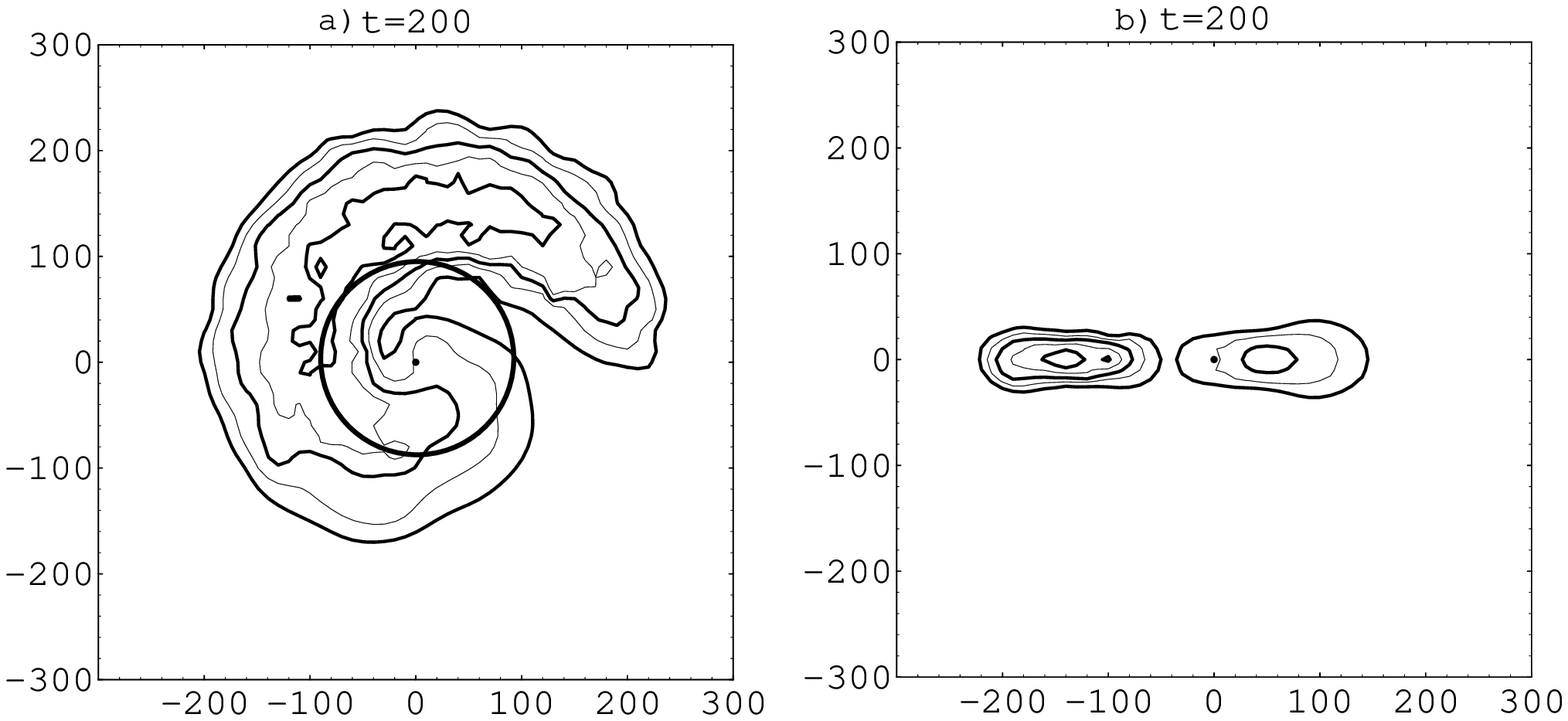,angle=0,clip=}
\caption{Density contours after the disruption of the polytrope for
the dynamical run using the Paczy\'{n}ski-Wiita potential with $q=0.1$
in the equatorial plane (a), and in the meridional plane containing
the black hole (b). All contours are logarithmic and equally spaced
every 0.5 dex. The lowest contour is at $\log{\rho}=-10$ (in the units
defined in equation~[\ref{eq:defrhounit}]), and bold contours are
plotted at $\log{\rho}=-10,-9,-8$. The thick black line in panel (a)
divides matter which is bound to the black hole (small black disk at
center) with that which is on outbound trajectories.}
\label{rhocontoursq01pw}
\end{figure*}

Since gravitational interactions are stronger with the modified form
of the gravitational potential, the overall encounter is more
violent. The neutron star is tidally disrupted into a long tidal tail
in a way similar to that exhibited in run E. The accretion episode is
very brief, with a peak accretion rate onto the black hole of
$\dot{M}_{max}=0.7$, equivalent to 8.5~M$_{\odot}$/ms and thus
substantially higher than that for run E (see Table~\ref{disks}).

\begin{figure}
\psfig{width=85mm,file=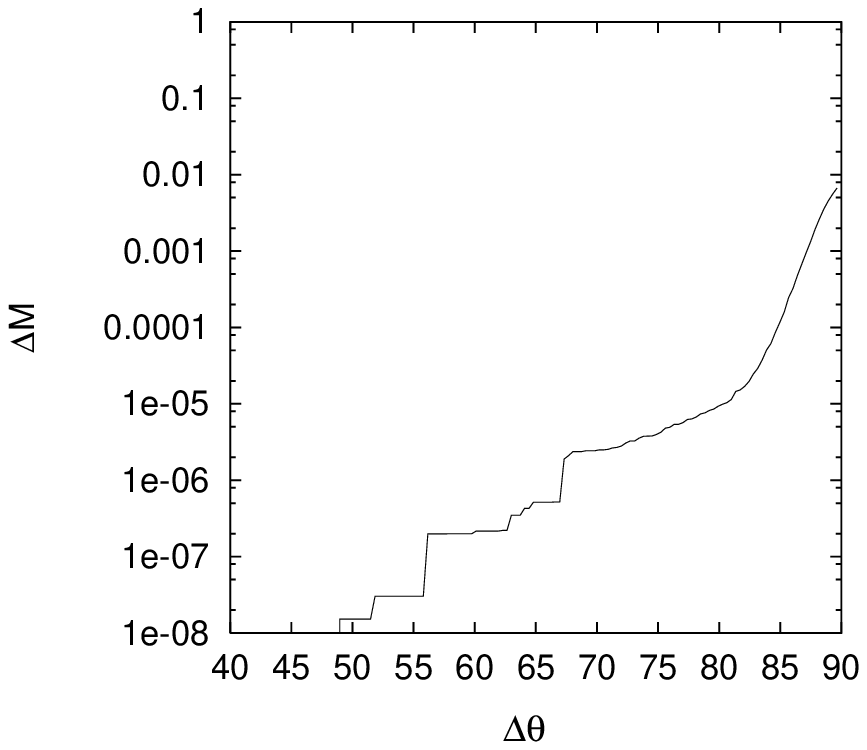,angle=0,clip=}
\caption{Enclosed mass as a function of half--angle $\Delta \theta$
(measured from the rotation axis in degrees) for the run using the
Paczy\'{n}ski--Wiita potential at $t=t_{f}$. The mass resolution is on
the order of $10^{-6}$ (correspoding to $1.4 \times 10^{-6}$
M$_{\odot}$) for an opening angle of approximately 65 degrees (compare
with Figure~\ref{mtheta}, and note that the x--axis is {\em not}
logarithmic in this case).}
\label{mthetapw}
\end{figure}

We followed the dynamical evolution from $t=0$ to $t_{f}=200$, and
show final density contour plots in the orbital and meridional plane
containing the black hole in Figure~\ref{rhocontoursq01pw}.  By the
end of the simulation, the fluid has not formed a quasi--static
accretion structure around the black hole as for the Newtonian runs,
and 99.2\% of the initial neutron star mass has been accreted
($M_{acc}=0.992$). The thick black line in
Figure~\ref{rhocontoursq01pw}a divides material that is bound to the
black hole from that which is on outbound trajectories (see
Figure~\ref{ejected} for a comparison with runs D and E). Overall, a
smaller amount of mass is left over after the initial episode of heavy
mass transfer (approximately an order of magnitude less than for run
E), but a larger fraction ($M_{ejected}=6.8 \times 10^{-3}$,
equivalent to $9.6 \times 10^{-3}$M$_{\odot}$) may be dynamically
ejected from the system. The region above and below the black hole is
devoid of matter to an even greater extent than for the Newtonian case
as can be seen in Figure~\ref{mthetapw},
where we plot the enclosed mass $\Delta M$ as a function of the half
angle $\Delta \theta$ of a cone directly above (and below) the black
hole at $t=t_{f}$, as in Figure~\ref{mtheta}. Thus in this scenario
the production of a relativistic fireball that could give rise to a
gamma--ray burst would require an even more modest degree of beaming
in order to avoid baryon contamination.

\label{lastpage} 

\end{document}